\documentclass[10pt]{iopart}

\usepackage{graphicx,siunitx,array,makecell,hhline,multirow,url,hyphenat,eqnarray} 
\DeclareSIUnit\px{px}
\DeclareSIUnit\frame{frame}
\newcolumntype{M}[1]{>{\centering\arraybackslash}m{#1}}
\graphicspath{{../Proto-Rig_Measurements/}}

\begin{document}

\title{Towards quantitative small-scale thermal imaging}
\author{J L McMillan, A Whittam, M Rokosz and R C Simpson}
\address{National Physical Laboratory, Teddington, TW11 0LW. UK}
\ead{jamie.mcmillan@npl.co.uk}

\begin{abstract}
Quantitative thermal imaging has the potential of reliable temperature measurement across an entire field-of-view. This non-invasive technique has applications in aerospace, manufacturing and process control. However, robust temperature measurement on the sub-millimetre (\SI{30}{\micro\metre}) length scale has yet to be demonstrated. Here, the temperature performance and size-of-source (source size) effect of a \SIrange{3}{5}{\micro\metre} thermal imaging system have been assessed. In addition a technique of quantifying thermal imager non-uniformity is described. An uncertainty budget is constructed, which describes a measurement uncertainty of \SI{640}{\milli\kelvin} \((k=2)\) for a target with a size of \SI{10}{\milli\metre}. The results of this study provide a foundation for developing the capability for confident quantitative sub-millimetre thermal imaging.

\end{abstract}


\maketitle
\ioptwocol

%--------------------------------------------------------
%	Introduction
%--------------------------------------------------------

\section{Introduction}\label{sec:intro}
\subsection{Small-Scale Thermal Imaging and its Applications}\label{subsec:lit_ss_app}
Thermal imaging in the sub-millimetre scale has been employed in diverse fields, spanning integrated circuits and living cells to the measurement of atomic interactions \cite{temp_map_gan_led,quan_ir_therm_ss_oop_def,nano_therm_cell}. Thermal imaging offers a technique to non-destructively measure temperature across an entire field-of-view. Compared to contact thermometry, it offers non-invasive temperature assessment for the region of interest.  

Although temperature measurement at the sub-millimetre scale has been performed, measurements have not fully evaluated the optical and material effects such as: size-of-source, non-uniformity and emissivity. From a metrological perspective, these need to be controlled and accounted for. Novel alternatives exist for nanometre temperature measurement \cite{nano_therm_cell}, however these methods cannot demonstrate traceability to the International Temperature Scale of 1990 (ITS-90). This is problematic for applications where robust, internationally accepted temperatures with defined uncertainties are required.

\subsection{Scope of Current Capabilities}\label{subsec:scope}
The purpose of this research was to robustly investigate the uncertainty of temperature measurement at the sub-millimetre scale using thermal imagers. Calibration of radiation thermometers is performed worldwide using agreed procedures (such as those detailed in \cite{cct_wg5_uncert_budg}); these are based on the use of black body reference sources using a heat-pipe (or equivalent) to minimise the thermal gradient along a long cylindrical cavity-type black body.

For sub-millimetre scale measurements, tailored lenses are used with thermal imaging systems to decrease the focal length and observe a smaller field-of-view.

\begin{figure}[t]
\centering
\includegraphics[width=0.45\textwidth,keepaspectratio]{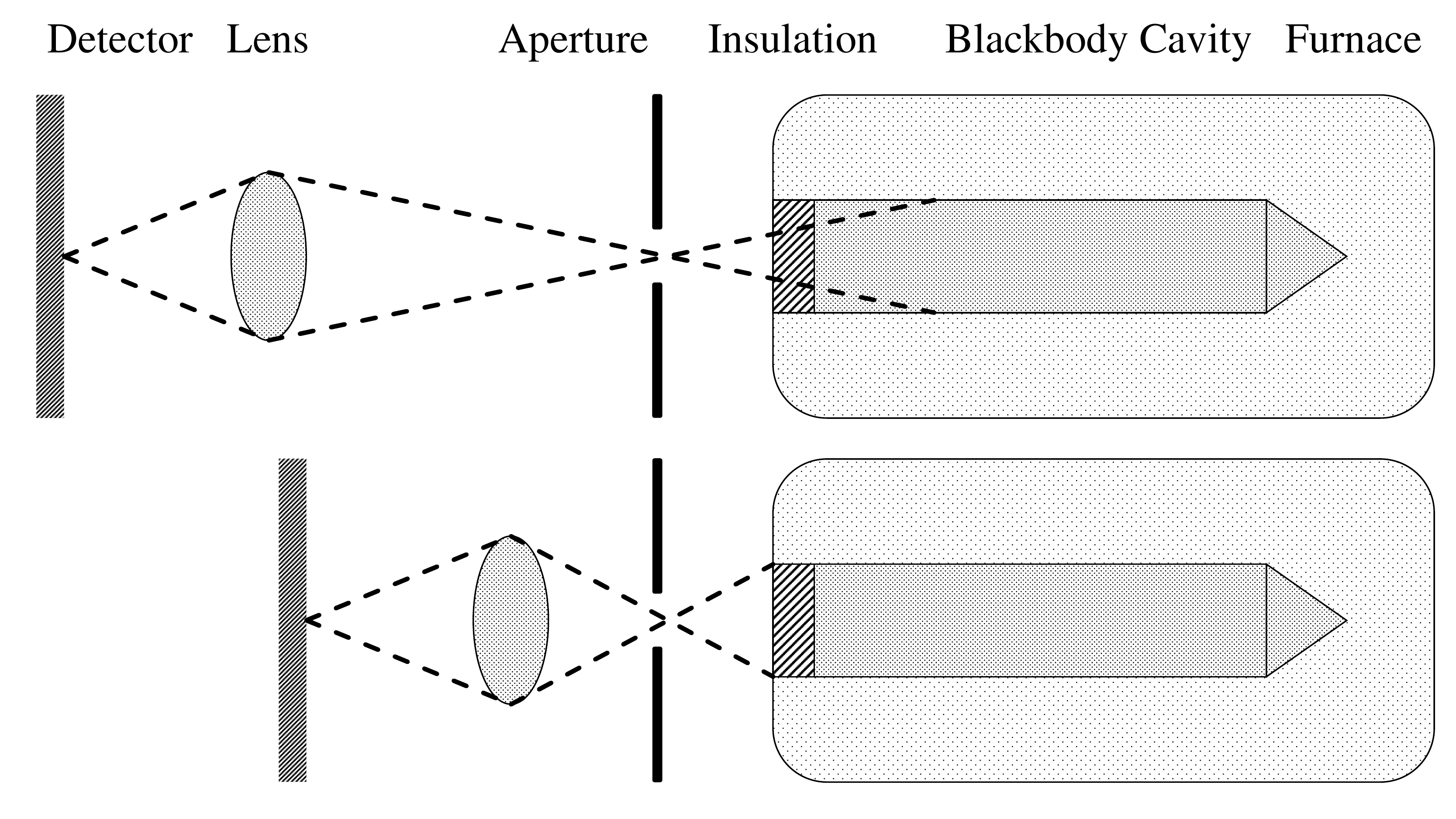}
\caption{The optics of the thermal imager and black body cavity. Typically (top) the radiation is collected onto the detector from the nominally uniform and freely radiating black body by focusing on the aperture. Using the macro lens (bottom), the reduced focal length leads to the detector measuring radiation emitted from the non-uniform side walls (due to convection near the aperture) and insulation of the cavity.}
\label{fig:macro_lens_cavity}
\end{figure}

When calibrating thermal imagers suitable for small-scale temperature measurement an issue arises from the current setup (visualised in figure \ref{fig:macro_lens_cavity}) caused by the elongated nature of the cavity. For a typical lens and thermal imager focused on the aperture plate, whilst the radiation is not collected from a single point on the back wall, the radiance temperature is well controlled within this region. Instruments can either be focused on the aperture plate assuming the cavity is uniform in temperature along its length, or if this cannot be guaranteed they can be focused on the back wall. However, using thermal imagers with the macro lens, the reduced focal length results in radiation being detected from insulation at the entrance to the cavity and its walls. Additionally, when focus on the back wall is required, this becomes difficult with a short focal length.

\subsection{Flat-Plate Black Bodies}\label{subsec:lit_bb}
Flat-plate black bodies have been available for over a decade and in principle would be well suited to the temperature calibration of sub-millimetre scale artifacts \cite{spec_calib_bb}. As depicted in figure \ref{fig:macro_lens_cavity}, the field-of-view when employing typical black bodies for a thermal imager with a focal length of a few millimetres, would measure a non-thermally-uniform region of the cavity. Attempts to mitigate this by designing cavities with larger apertures, exhibit less control of the temperature gradients along the cavity length due to convection. An uncertainty of \SI{0.2}{\kelvin} \((k=2)\) in the reduced temperature range of \SIrange{280}{340}{\kelvin} has been reported for a design larger than \SI{250}{\milli\metre} \cite{lar_ap_bb_bath,cons_char_lar_ap_bb}.

Flat-plate black bodies are an alternative to cavities, and consist of a temperature controlled surface with a nominally uniform emissivity surface. Flat-plate calibrators exhibit lower emissivites than cavity-like black bodies due to their geometry; but their large surface area enables the calibration of a wider range of instruments.

This paper aims to gauge the design requirements that a calibration facility would need to overcome the technical challenges encountered calibrating thermal imagers tailored for small-scale temperature measurement. The initial concept was to develop a flat-plate black body based on the use of vertically aligned carbon nanotubes, due to recent publications underlining their impressive properties as a black body radiator \cite{bb_abs_va_swcn,hot_nano_mwcn_2000,bb_rad_res_mwcn}. However, manufacturing a sample of these nanotubes costs in the region of \textsterling10,000 (technique based on \cite{grow_cnt_comp_ic}) and so instead a preliminary experimental setup was performed using a Nextel coated plate.

\subsection{Nextel-Black Coating}\label{subsec:lit_nextel}
A conventional research grade high-emissivity coating is Nextel-Velvet coating 811-21 manufactured by Mankiewicz Gebr. \& Co. (Hamburg, Germany). Its thermal and optical properties have been well characterised. Early tests on its thermal conductivity and total emissivity showed values of \SI{0.195}{\watt\per\metre\per\kelvin} at \SI{345}{\kelvin}, and total emissivity remained above \num{0.95} for incident angles less than \SI{50}{\degree} and wavelengths between \SI{5}{\micro\metre} and \SI{30}{\micro\metre} \cite{therm_cond_nextel}.

A comparison between common black coatings included: a zinc plate, NPL Super Black, Pyromark, Nextel and commercial matt black car paint. The Nextel sample had an emissivity above \num{0.96} over the \SIrange{2}{5}{\micro\metre} spectral range. It typically had a higher emissivity, so better suitability as a characterisation source, than commercial flat-plate calibrators \cite{comm_bl_coat_ref_age}.

\vspace{1em}
The following manuscript details the experimental methods employed and their results. It consists of an overview of the experimental setup, a detailed description of the measurements taken, the results and an uncertainty budget.

%--------------------------------------------------------
%	Method
%--------------------------------------------------------

\section{\bf Method} \label{sec:method}
\subsection{Experimental Setup} \label{subsec:setup}
The experimental setup consisted of an aluminium block (\num{90}\( \times \)\num{90}\( \times \)\SI{6}{\milli\metre}) coated with Nextel paint (referred to as the sample throughout this paper), resting on a resistive heating mat, both compressed between two aluminium plates. A single coat of Nextel paint was applied directly to polished aluminium, using no primer, by spray gun. This was mounted upon translation and rotational stages, separated by \SI{6}{\milli\metre} of insulation. A \SI{100}{\ohm} platinum resistance thermometer (PRT) controlling the heater mat was located below the bottom plate, and an identical reference PRT to measure the surface temperature was adhered directly to the sample, \SI{30}{\milli\metre} from the centre of the measurement region. The entire mount was contained within a thermally insulating enclosure. In order to observe the sample surface the top aluminium plate (enclosing the sample and heater) and lid of the enclosure had a \SI{40}{\milli\metre} aperture. Above the enclosure, a Cedip Silver thermal imager (\SIrange{3}{5}{\micro\metre}) was rigidly mounted to view the sample surface, at a separation of \SI{16.5}{\milli\metre} between thermal imager lens and top of sample (unless otherwise stated).

The reference PRT was measured using an ASL F700 resistance ratio bridge. The standard resistor temperature was monitored throughout the entire measurement period; the temperature remained at \SI{293.6+-0.1}{\kelvin} (throughout this paper the notation `mean (standard deviation)' is used). As the temperature change was minimal, no correction was applied for standard resistor drift. The internal camera housing temperature of the thermal imager remained at \SI{308.88+-0.32}{\kelvin} during the measurements.
% Camera temperature from using:
% cut -f 8 ~/CNT/{Non-Uniformity,Temperature,SSE}/**/*nextel_v2_rad_*_*.csv | sed '/^[a-z][A-Z]/d;/Camera/d' | sort -n | stats

Radiance temperature was measured using the manufacturer supplied software, emissivity was set to \num{1.00} and the integration time was fixed throughout the measurements. A \SI{46}{\px} (pixel) circular region (approximately \SI{220}{\micro\metre}, derived from the value in section \ref{subsec:nu_meth}) located at nominally the centre of the flat-plate, referred to as the central region, was recorded.

\subsection{Temperature and Stability}\label{subsec:temp_meth}
A series of characterisation tests on the sample-imaging system were performed in order to gauge its temperature stability and document its measurement uncertainty.

To assess the behaviour of the setup over a range of temperatures, the difference between thermal imager temperature and the reference PRT was measured. The heating element was programmed to a range of temperatures; once stable (close to reference PRT resolution) a \SI{2}{\minute} set of data was recorded. Additionally, the stability of the system over a period of approximately \SI{2}{\hour} was monitored at nominally \SI{320}{\kelvin}.

\subsection{Size-of-Source Effect}\label{subsec:sse_meth}
The size-of-source effect (SSE), was assessed using the direct method \cite{corr_rad_therm_sse}. A set of black-coated brass apertures varying from \SIrange{0.75}{25}{\milli\metre} were used. These lay on top of the enclosure and rested against a guide to ensure geometrically repeatable placement. To understand the effect distance between thermal imager and source has on the measurements, as well as any relation between the focal range for the thermal imager and temperature, four sets of data were collected from \SIrange{15}{19}{\milli\metre} separation between the front of the lens and the aperture. This spans \(63\,\%\) of the entire focal range of the thermal imager. At each of these four positions, three sets of data were collected. 

Both PRT and radiance temperatures were sampled from the data, and radiance was corrected for any drift using the PRT data. The SSE was then calculated by first converting to quasi-spectral radiance (based on Planck's Law), then using the equations below \cite{sse_temp_uncert_low_temp}. Here, radiance is \(L\), \(c_2=\SI{0.014388}{\metre\kelvin}\) is the second radiation constant, \(\lambda\) is the wavelength (assumed to be \SI{4}{\micro\metre}) and \(T\) the temperature in kelvin. The calculated SSE values for the four distances were then averaged between the three repeats.

\begin{center}
\(L(T)=\frac{1}{exp(\frac{c_2}{\lambda T})-1}\) \hspace{2em} \(SSE(L)=\frac{L(T)}{L_{max}(T)}\)
\end{center}

\subsection{Non-Uniformity}\label{subsec:nu_meth}
\begin{figure}[t]
\centering
\includegraphics[width=0.45\textwidth,trim={0.30cm 16.5cm 14.5cm 1.4cm},clip,keepaspectratio]{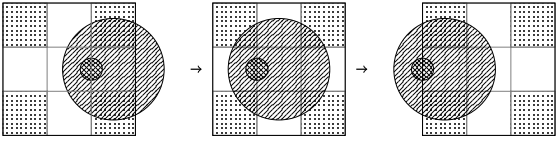}
	\caption{An example of three frames to be aligned and summed. Here the (square) \num{3}\( \times \)\num{3} resolution (non-uniform) thermal imager is being translated over the stationary (non-uniform and circular) sample from left to right, and the arrays captured are \(\underline{\underline{A}}\), \(\underline{\underline{B}}\) and \(\underline{\underline{C}}\), respectively. Beyond the radiance field at each pixel being depicted, each array additionally describes both imager and sample non-uniformity.}
\label{fig:lin_trans_frames}
\end{figure}

Here the process used to measure the non-uniformity of the detector is described. This does not make any assumption on the uniformity of either the detector or sample. A linear translation of a sample across the thermal imager focal plane array can be considered a sequence of temperature arrays, or frames. By aligning a number of these frames on a central frame and averaging them, the component of the thermal imager non-uniformity can be identified. This process is visualised in figure \ref{fig:lin_trans_frames}, depicting the three arrays \(\underline{\underline{A}}\), \(\underline{\underline{B}}\) and \(\underline{\underline{C}}\), respectively. The notation \(a_{ij}\) refers to the \(i^{th}\) row and \(j^{th}\) column of matrix \(A\). 

\begin{center}
\( \underline{\underline{A}} = \left[ \begin{array}{ccc}
a_{11} & a_{12} & a_{13} \\  
a_{21} & a_{22} & a_{23} \\  
a_{31} & a_{32} & a_{33} \end{array} \right] \) \;\;
\( \underline{\underline{B}} = \left[ \begin{array}{ccc}
b_{11} & b_{12} & b_{13} \\  
b_{21} & b_{22} & b_{23} \\  
b_{31} & b_{32} & b_{33} \end{array} \right] \)
\end{center}
\begin{center}
\( \underline{\underline{C}} = \left[ \begin{array}{ccc}
c_{11} & c_{12} & c_{13} \\  
c_{21} & c_{22} & c_{23} \\  
c_{31} & c_{32} & c_{33} \end{array} \right] \)
\end{center}

The three frames in the sequence are aligned upon frame \(\underline{\underline{B}}\), by shifting \(\underline{\underline{A}}\) and \(\underline{\underline{C}}\) right \(1 \cdot \delta\) and left \(-1 \cdot \delta\) columns, respectively, where \(\delta\) is the pixel shift between successive frames, and in this case \(\delta=\SI{1}{\px\per\frame}\). The \(k^{th}\) frame from the central frame is shifted by \(k \cdot \delta\) columns. This summed array \(\underline{\underline{S}}\) is equivalent to the following matrix addition.

\(\underline{\underline{S}} = \left[ \begin{array}{ccc}
0 & a_{11} & a_{12} \\  
0 & a_{21} & a_{22} \\  
0 & a_{31} & a_{32} \end{array} \right] +
\left[ \begin{array}{ccc}
b_{11} & b_{12} & b_{13} \\  
b_{21} & b_{22} & b_{23} \\  
b_{31} & b_{32} & b_{33} \end{array} \right] +
\left[ \begin{array}{ccc}
c_{12} & c_{13} & 0 \\  
c_{22} & c_{23} & 0 \\  
c_{32} & c_{33} & 0 \end{array} \right] =
\left[ \begin{array}{ccc}
b_{11} + c_{12} & a_{11} + b_{12} + c_{13} & a_{12} + b_{13} \\  
b_{21} + c_{22} & a_{21} + b_{22} + c_{23} & a_{22} + b_{23} \\  
b_{31} + c_{32} & a_{31} + b_{32} + c_{33} & a_{32} + b_{33} \end{array} \right] \)

\vspace{1em}
To account for the fringe-pattern between different columns, due to the number of frames summed, a normalisation is required. By assuming in this example \(\overline{a_{ij}}\approx \overline{b_{ij}}\approx \overline{c_{ij}}\approx 1 = \alpha\), then;

\begin{center}
\( \underline{\underline{S}} \approx \left[ \begin{array}{ccc}
2 & 3 & 2 \\  
2 & 3 & 2 \\  
2 & 3 & 2 \end{array} \right] \)
\end{center}

\noindent The normalisation follows:

\begin{center}
\(n_{ij} = \frac{s_{ij}}{Int\left(\frac{\overline{\underline{S}_{j}}}{\alpha}\right)}\)
\end{center}

Here the average value of column \(\underline{S}_{j}\) is divided by a scaling factor \(\alpha\) and the value is rounded to an integer. This indicates the number of frames comprising the \(j^{th}\) column and normalises the element \(s_{ij}\), resulting in a translation-corrected array \(\underline{\underline{N}}\).

\begin{center}
\( \underline{\underline{N}} = \left[ \begin{array}{ccc}
n_{11} & n_{12} & n_{13} \\  
n_{21} & n_{22} & n_{23} \\  
n_{31} & n_{32} & n_{33} \end{array} \right] \approx
\left[ \begin{array}{ccc}
1 & 1 & 1 \\  
1 & 1 & 1 \\  
1 & 1 & 1 \end{array} \right] \)
\end{center}

This interprets the detector non-uniformity as noise and the sample non-uniformity as signal; for each pixel, the signal persists in each frame whilst the noise is distributed across a number of pixels. The resulting signal-to-noise ratio of the sample to detector non-uniformity will be increased and this can then be used to isolate the detector non-uniformity. 

The values \(\delta\) and \(\alpha\) for the measurement data collected were estimated from the data itself. \(\alpha\) was calculated from the average of all data points within the central region during the non-uniformity measurements. 

\(\delta\) was derived semi-subjectively; the alignment process was completed for values of \(\delta\) between \SIrange{8.60}{8.70}{\px\per\frame} and the mid-point at which the frames were over- or under-shifted was selected, such that \(\delta=\SI{8.67+-0.005}{\px\per\frame}\). The displacement per frame was \SI{250+-5}{\micro\metre\per\frame}, therefore the size-per-pixel was \SI{29+-7}{\micro\metre\per\px}. The field-of-view at a distance of \SI{16.5}{\milli\metre} was \num{9.28}\( \times \)\SI{7.42}{\milli\metre}.

The alignment process governed by \(\delta\) is limited to integer steps of \SI{1}{\px}. To gauge the uncertainty incurred from this process a mean neighbour test was employed. For each of the original frames, each pixel was located and its surrounding box of nominally eight pixels were compared and the mean recorded. For each of the three repeats, the mean value of each frame was recorded.

%--------------------------------------------------------
%	Temperature & Stability
%--------------------------------------------------------

\section{\bf Results}\label{sec:res}
\subsection{Temperature and Stability}\label{subsec:temp_stab}
\begin{figure}[t]
\centering
\includegraphics[width=0.45\textwidth,keepaspectratio]{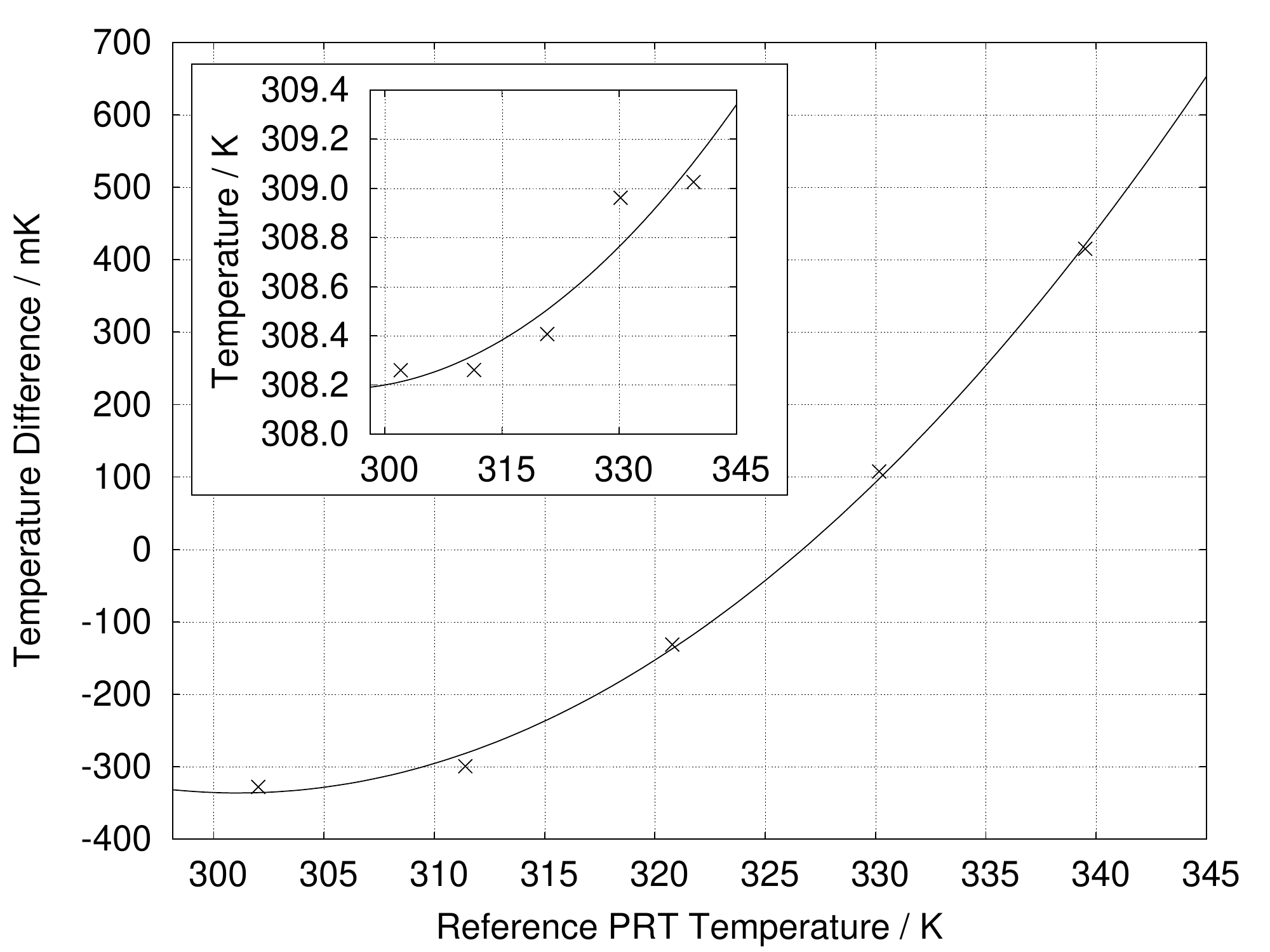}
\caption{The difference between the PRT temperature and radiance temperature (\(T_{PRT}-T_{Rad}\)) over a range of temperatures. Inset depicts the internal camera housing temperature over the same range.}
\label{fig:temp}
\end{figure}

The temperature difference between the PRT and thermal imager (\(T_{PRT} - T_{Rad}\)) over an increasing temperature range is depicted in figure \ref{fig:temp}; the temperature was limited to this range by the heating element. Within the measurement range, the measured difference remained within \SI{+-0.5}{\kelvin}.

A correlation between sample radiance temperature and internal camera housing temperature (figure \ref{fig:temp} inset) indicates that the close proximity of the thermal imager to the sample (\SI{16.5}{\milli\metre}) is likely to be a component of concern in developing the technique and expanding the temperature range.

More directly affected by the close proximity to the sample is the heating of the lens. This will alter the optical behaviour of the imaging system due to the temperature dependence of refractive index; this will be a significant source of uncertainty at higher temperatures.

The standard deviation of the PRT (attached to the sample surface) over \SI{2}{\hour} was \SI{7}{\milli\kelvin}, and was \SI{30}{\milli\kelvin} for the temperature measurement of the thermal imager, both at \SI{320.67}{\kelvin}. These are close to the resolution limits, \SI{1}{\milli\kelvin} and \SI{10}{\milli\kelvin}, respectively.

%--------------------------------------------------------
%	Size-of-Source Effect
%--------------------------------------------------------

\subsection{Size-of-Source Effect}\label{subsec:sse}
\begin{figure}[t]
\centering
\includegraphics[width=0.45\textwidth,keepaspectratio]{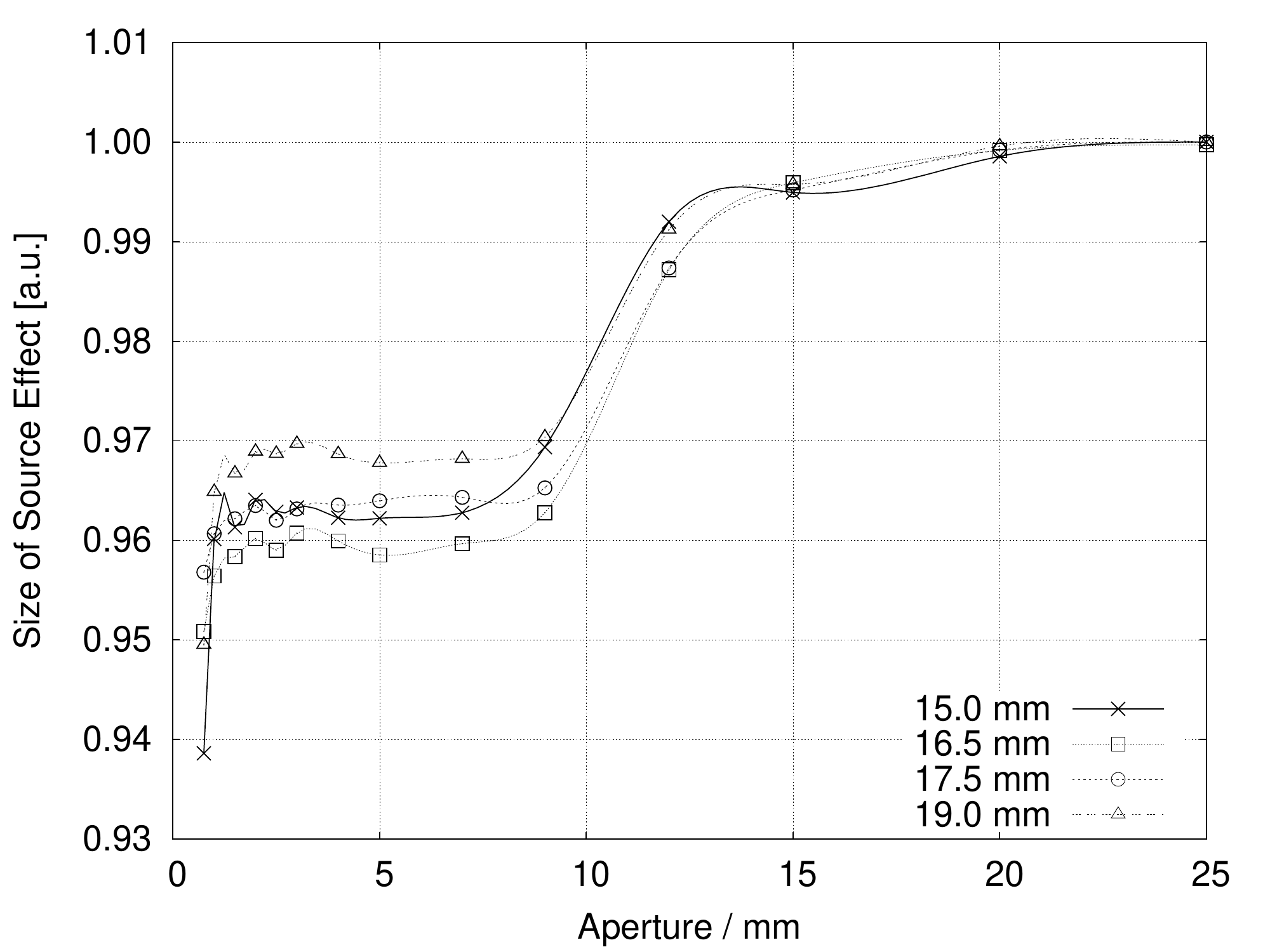}
\caption{The SSE values at each separation. Key indicates the difference values of separation between lens and sample surface. The data was consistent between runs, it appears to plateau at two values. There was no correlation between distance and SSE.}
\label{fig:sse}
\end{figure}

The size-of-source effect observed in typical radiation thermometers displays a sharply increasing value at lower aperture sizes and a plateau at larger aperture sizes. For the four separations assessed here, each follow this behaviour below the \SI{10}{\milli\metre} aperture. However, above this value there appears a second plateau occurring beyond \SI{15}{\milli\metre} (figure \ref{fig:sse}). The second plateau occurs beyond the field-of-view (nominally \num{9}\( \times \)\SI{7}{\milli\metre}) of the thermal imager.

These measurements were taken with the thermal imager focused on the aperture plates at separations of \SIrange{15.0}{19.0}{\milli\metre}, and correspondingly \SIrange{32.5}{36.5}{\milli\metre} between the thermal imager and sample.

%--------------------------------------------------------
%	Non-Uniformity
%--------------------------------------------------------

\subsection{Non-Uniformity}\label{subsec:nu}
\begin{figure}[t]
\centering
\includegraphics[trim={4.2cm 4.3cm 3.0cm 5.5cm},width=0.45\textwidth,keepaspectratio]{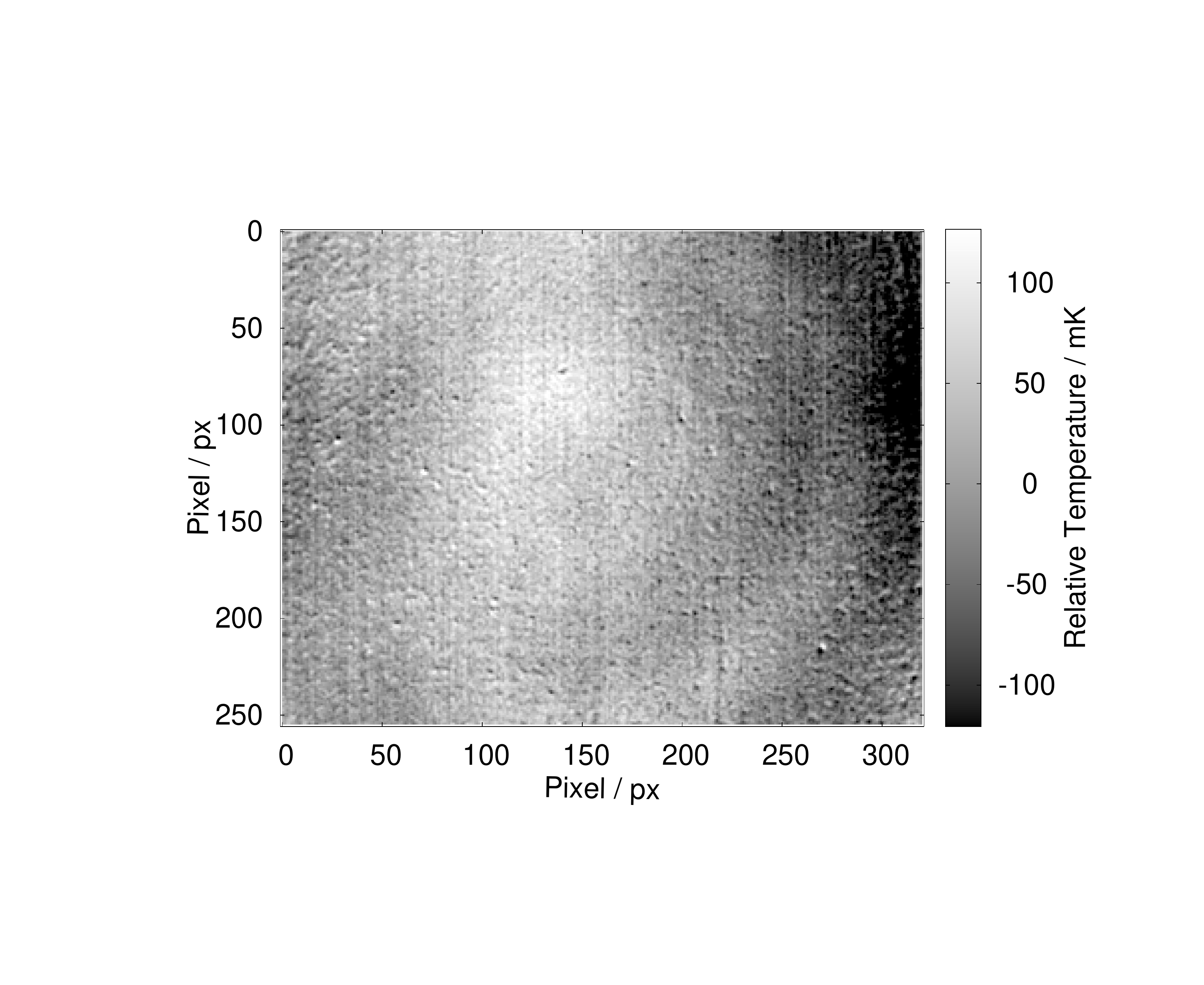}
\caption{Non-uniformity across the thermal imager, the standard deviation of the temperature across this image is \SI{62}{\milli\kelvin}.}
\label{fig:imager_nu}
\end{figure}

A normalised translation-aligned array (equivalent to \(\underline{\underline{N}}\)) ideally depicts the central frame of the sequence, free from sample non-uniformity and, due to the frame-by-frame PRT drift correction, free from drift. To then visualise the thermal imager non-uniformity, the original frame is subtracted from the aligned frame (i.e., \((T_{sample}+u_{imager})-T_{sample}\equiv u_{imager}\)). An example of this is visualised in figure \ref{fig:imager_nu}. For the three repeat sets of measurement, the average temperature and standard deviation across the entire image were: \SI{12+-65}{\milli\kelvin}, \SI{3+-62}{\milli\kelvin} and \SI{-7+-64}{\milli\kelvin}. More critical than the global analysis of these arrays was the structure within the array itself, such that a correction to future measurements may need to be applied.

\begin{figure}[t]
\centering
\includegraphics[trim={0cm 0cm 0cm 0.5cm},width=0.45\textwidth,keepaspectratio]{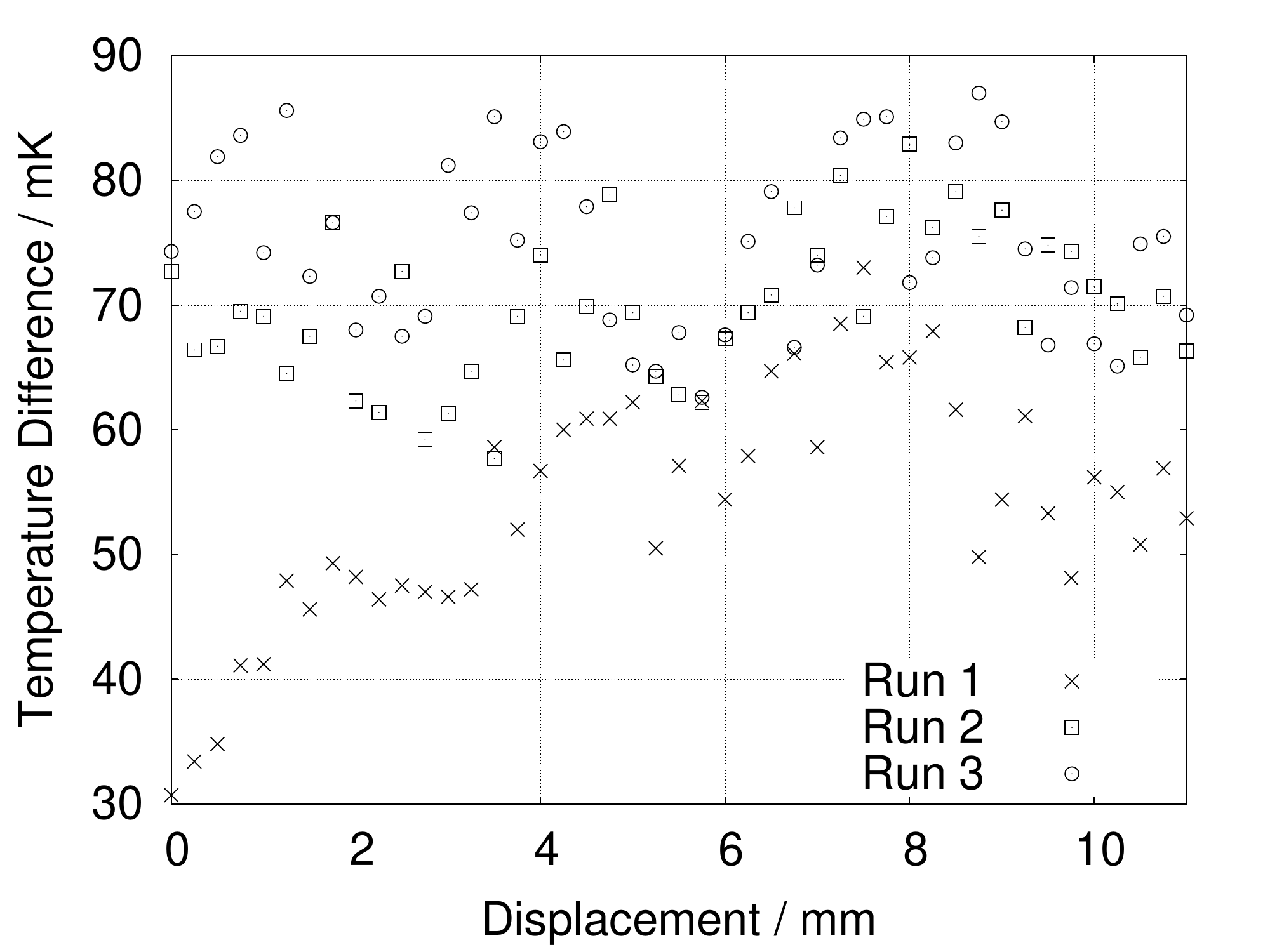}
\caption{The mean neighbour for each frame in the translation sequence. The uncertainty introduced from the non-uniformity correction technique, of \SI{90}{\milli\kelvin}.}
\label{fig:neigh_comp}
\end{figure}

The uncertainty introduced through this correction technique can be evaluated by calculating the incurred error by misaligning by \SI{1}{\px}. For each frame in the sequence, the mean neighbour was evaluated for each pixel. This mean neighbour was averaged across the entire sequence, for each of the three repeats; this is plotted in figure \ref{fig:neigh_comp}.

%--------------------------------------------------------
%	Surface Topology
%--------------------------------------------------------

\subsection{Surface Topology}\label{subsec:surf_top}
\begin{figure}[t]
\centering
\includegraphics[trim={4.2cm 4.3cm 3.0cm 5.5cm},width=0.45\textwidth,keepaspectratio]{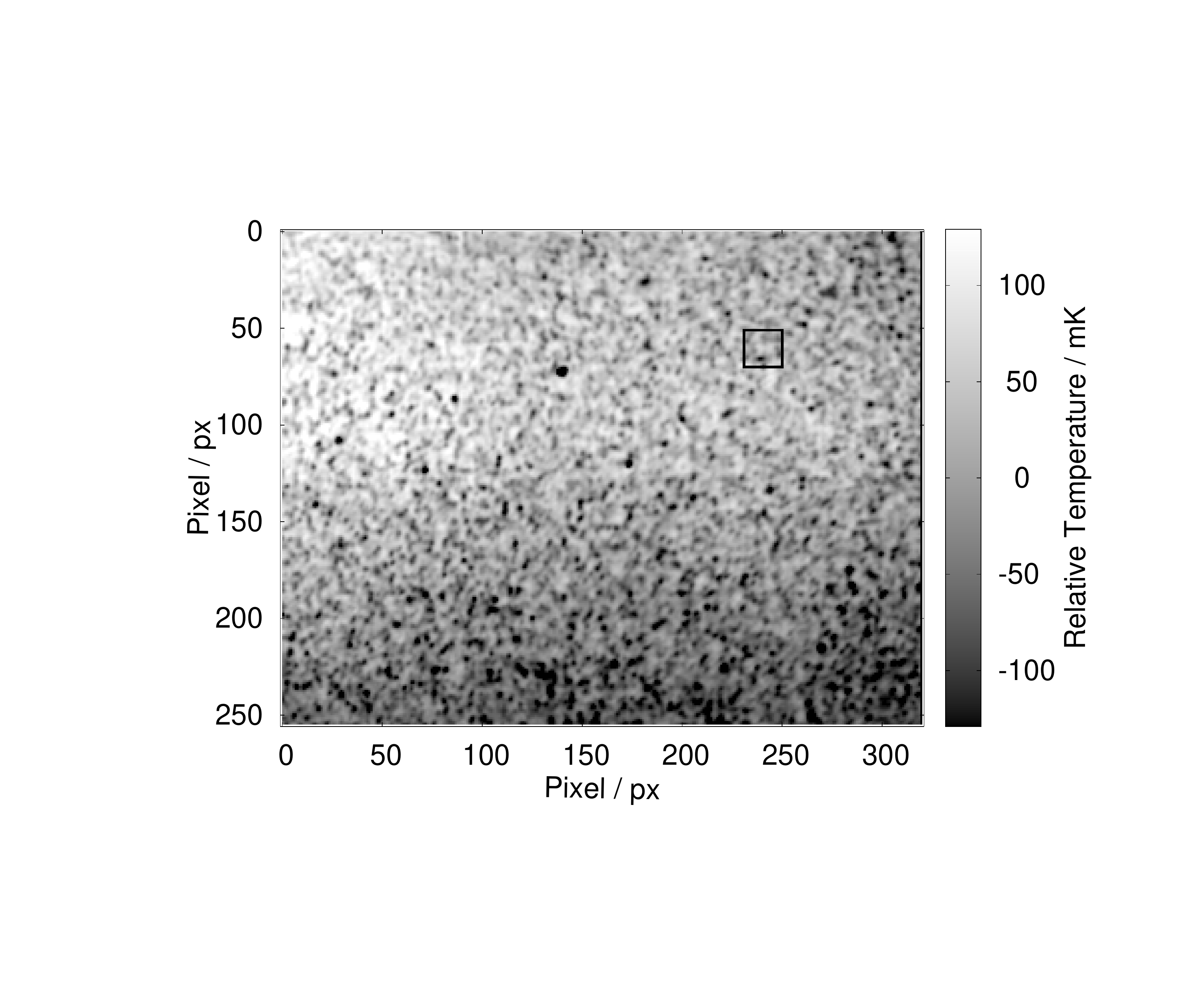}
\caption{Temperature non-uniformity across the sample relative to the average, the average temperature across this image is \SI{320.67+-0.06}{\kelvin}. Highlighted region is a \SI{20}{\px} square, evaluated in section \ref{sec:disc}.}
\label{fig:sample_nu}
\end{figure}

When observing the thermal imager non-uniformity-corrected sample, localised temperature perturbations on the surface of the flat-plate can be seen (figure \ref{fig:sample_nu}). The average temperature across this array is \SI{320.67}{\kelvin} and the standard deviation across it is \SI{60}{\milli\kelvin}.
% cat ~/CNT/Non-Uniformity/Nextel_v2/Run_1/SubRun_2/Altair/Temperatures/nu_50degC_nextel_v2_20.50mm_1_2_camera_nu_and_drift_corrected.csv | tr '\t' '\n' | stats

To understand the effect on temperature physical roughness presents, an assessment of the surface topology was performed. An Alicona 3D optical surface measurement system was used as a profiler, it had a \SI{7}{\micro\metre} resolution and so has a higher spatial resolution than the thermal imager. No visual-infrared reference point was used, hence a comparison between the data in figure \ref{fig:sample_nu} and a region nominally the size of the thermal imager field-of-view in the topology data was performed.

No direct correlation between the temperature perturbations and surface height was observed. These perturbations are unlikely a true temperature difference over a \SI{30}{\micro\metre} region, and instead originate from some radiance temperature variation. A detailed investigation of the specific emissivity implications small height fluctuations create may yield a clearer understanding.

%--------------------------------------------------------
%	Uncertainties
%--------------------------------------------------------

\section{\bf Uncertainties}\label{sec:uncert}
The uncertainty evaluation of the experiment is detailed in Table \ref{tab:uncert} for both the sample and thermal imager. It assumes a temperature of \SI{320.67}{\kelvin} and is valid for the full field-of-view. This budget was assembled in accordance with \cite{cct_wg5_uncert_budg} and the definition of each source is described below.

% This figure belongs to subsec:surf_top but is here to place it on the top of the page.

\begin{table*}[ht]
\centering
\begin{tabular}{ !{\vrule width 2pt}M{1.75cm}|M{3.85cm}||M{1.85cm}|M{1.85cm}|M{1.85cm}|M{1.85cm}!{\vrule width 2pt} }
\Xhline{2pt}
\multicolumn{6}{!{\vrule width 2pt}c!{\vrule width 2pt}}{Uncertainty Budget}  \\
\Xhline{2pt}
Instrument & \multicolumn{1}{c||}{Source} & u/\SI{}{\milli\kelvin} & Distribution & Divisor & U/\SI{}{\milli\kelvin} \\
\hline
\hline
\multirow{4}{*}{Sample} & Reference PRT & 13 & N & 1 & 13 \\
\cline{2-6}
 & Emissivity & 208 & N & 1 & 208 \\
\cline{2-6}
 & Reflected Radiation & 199 & N & 1 & 199 \\
\cline{2-6}
 & Uniformity & 60 & N & 1 & 60 \\
\hline
\hline
\multirow{6}{*}{Imager} & Size-of-Source Effect & 113 & R & \(\sqrt{3}\) & 65 \\
\cline{2-6}
 & Internal Reference & 50 & N & 1 & 50 \\
\cline{2-6}
 & Atmospheric Absorption & 6 & N & 1 & 6 \\
\cline{2-6}
 & Noise & 10 & N & 1 & 10 \\
\cline{2-6}
 & Drift & 30 & N & 1 & 30 \\
\cline{2-6}
\cline{2-6}
 & Uniformity & 90 & N & 1 & 90 \\
\Xhline{2pt}
\multicolumn{6}{!{\vrule width 2pt}c!{\vrule width 2pt}}{Combined Uncertainty \SI{640}{\milli\kelvin} \((k=2)\)} \\
\Xhline{2pt}
\end{tabular}
\caption{Defines the components comprising the uncertainty budget within the small-scale experimental setup, the first half relate to the sample and the second describe the thermal imager sources. These uncertainties were assessed in accordance with \cite{cct_wg5_uncert_budg}, they are valid for a full field-of-view measurement at \SI{320.67}{\kelvin}. The combined uncertainty was \SI{640}{\milli\kelvin} \((k=2)\).}
\label{tab:uncert}
\end{table*}

\subsection{Black Body Sample}\label{subsec:uncert_samp}
Components related to the black body sample include: reference PRT, emissivity, reflected radiation and uniformity.

\begin{itemize}
\item {\bf Reference PRT} temperature gauges the stability and precision of the reference thermometer.
% For stability during NU measurements 0.0128 degC
% cut -f 3 ~/CNT/Non-Uniformity/**/*nextel_v2_?_?.csv | sed '/[a-zA-Z]/d' | sort -n | stats
% For stability during stability test 0.0069 degC
% cut -f 2 ~/CNT/Long-Term_Drift/Nextel_v2/Run_1/drift_nextel_v2_prt_plotted_1.csv | sed '/[a-zA-Z]/d' | stats
% For resolution, look at the divisions of the stability graph

\item {\bf Emissivity} accounts for the emissivity and uncertainty of Nextel measured within \cite{comm_bl_coat_ref_age}.

\item {\bf Reflected ambient radiation} arises from the non-unity emissivity, and as such was dependent on the emissivity and uncertainty of the Nextel coating. The ambient conditions were controlled within \SI{294+-2}{\kelvin} during the measurements.
% Ambient temperature used is 21 degC.

\item {\bf Sample non-uniformity} has been assessed using the thermal imager correction in figure \ref{fig:imager_nu} to correct the measurements then calculate the standard deviation across the surface.
% The standard deviation across camera-NU-corrected central frames 0.06K
% for i in {1..3}; do cat ~/CNT/Non-Uniformity/Nextel_v2/Run_1/SubRun_$i/Altair/Temperatures/nu_50degC_nextel_v2_20.50mm_1_?_camera_nu_and_drift_corrected.csv | sed 's/\t/\n/g;s/0.00//g' | sort -n | stats; done 
\end{itemize}

\subsection{Thermal Imager}\label{subsec:uncert_imager}
The components related to the thermal imager include: size-of-source, internal reference, atmospheric absorption, noise, drift and non-uniformity.

\begin{itemize}
\item The {\bf size-of-source} effect standard deviation of repeat values has been converted into kelvin for its uncertainty.

\item {\bf Internal reference} temperature of the thermal imager detector temperature stability, was found to be stable to within \SI{50}{\milli\kelvin} over a period of \SI{2}{\hour}.

\item {\bf Atmospheric absorption}, based on the assumptions in \cite{cct_wg5_uncert_budg} measures the magnitude of the effect as \(0.02\,\%\) reduction of signal over the separation of nominally \SI{20}{\milli\metre}.

\item {\bf Noise} and {\bf drift} of the thermal imager are evaluated as the resolution and stability of the thermal imager, respectively.
% For stability during NU measurements 0.0471 degC
% cut -f 4 ~/CNT/Non-Uniformity/**/*nextel_v2_?_?.csv | sed '/[a-zA-Z]/d' | sort -n | stats
% For stability during stability test 0.0273 degC
% cut -f 6 ~/CNT/Long-Term_Drift/Nextel_v2/Run_1/drift_nextel_v2_rad_plotted_1.csv | sed '/[a-zA-Z]/d' | stats
% For resolution, look at the divisions of the stability graph

\item {\bf Non-uniformity} across the thermal imager has been identified in figure \ref{fig:imager_nu}, the uncertainty this correction introduces was quantified through figure \ref{fig:neigh_comp}.
\end{itemize}

%--------------------------------------------------------
%	Discussion
%--------------------------------------------------------

\section{\bf Discussion}\label{sec:disc}
The size-of-source measurements underline a limitation on the measurement of temperature at the sub-millimetre range and values of SSE below \num{1.00} indicate higher susceptibility to sources external to the target region. The \SI{0.75}{\milli\metre} aperture suffered poor reproducibility, however this was likely caused by the aperture clipping the central region. The two-step plateau is touched upon within \cite{charac_rad_therm_sse}, it may be a result of internal reflections at the surface of the objective lens.

Given the accepted convention that SSE above \num{0.98} is adequate, these results indicate that any temperature measurement of a region within the field-of-view size is unsuitable. For example, given an SSE value of \num{0.94} and a temperature of \SI{320}{\kelvin} at the maximum aperture, this would produce a \SI{2}{\kelvin} correction. For quantitative temperature measurement, a refined aperture set with a finer step size and a higher proportion in the sub-millimetre region will have to be used. This would enable more reliable correction in the region of interest.

The approach to SSE assessment described in this paper follows standard practice applied to radiation thermometers. What is not investigated is the validity of an SSE correction on a pixel other than the central pixel. The radiance observed by each pixel will differ and hence the severity of uncertainty incurred through the SSE will vary for each pixel \cite{sse_corr_therm_im}. This will be explored in a later study.

Non-uniformity of the radiance temperature from the radiating surface becomes apparent when viewed by the magnified thermal imager; although not a large contributor to the total uncertainty, its origin is not fully understood. The profiling of the topology showed little correlation to the apparent temperature variations observed within figure \ref{fig:sample_nu}. The source could be small emissivity fluctuations across the surface resulting from physical peaks and troughs; additionally from the same structural features, variations in thermal conductivity may lead to apparent temperatures differences.

The largest contributor to the uncertainty budget was the sample and its components, due to the less than unity emissivity. These components would be reduced through the use of vertically aligned carbon nanotubes to give a sample with higher emissivity. For comparison, the black body cavities used at NPL have an uncertainty from \SIrange{0.05}{0.30}{\kelvin} \((k=2)\) between \SI{230}{\kelvin} and \SI{1270}{\kelvin} \cite{hq_bb_ir_therm_40_1000}.

The impact of the calibration tests performed can be conveyed through a comparison between a naive approach and the application of the measurements performed in this paper. The naive average temperature within the space can be measured, using solely manufacturer provided software to adjust target emissivity. The manufacturer quoted accuracy (\SI{+-2}{\kelvin} or \SI{+-2}{\percent} of reading) can be attributed to the value. Indicated by the square in figure \ref{fig:sample_nu} is a \SI{20}{\px} sided region, which approximates to a \SI{0.34}{\milli\metre\squared} target. By applying the thermal imager non-uniformity correction from figure \ref{fig:imager_nu} to the square region, correcting for an emissivity of \num{0.96}, an SSE value of \num{0.95} and correcting to the PRT measurement from figure \ref{fig:temp}, a calibrated measurement and the standard uncertainty can be calculated.

\begin{center}
\begin{tabular}{lcc}
\(T_{naive}\) & \(=\) & \num{321.75}\SI{+-2.00}{\kelvin}\\
 & & \\
\(T_{calib}\) & \(=\) & \num{323.29}\SI{+-0.64}{\kelvin}\\
\end{tabular}
\end{center}

\noindent Assuming \(T_{calib}\) corresponds to a true value then considering the uncertainties, \(T_{naive}\) falls within the true value but the calibrated result has a lower uncertainty by a factor of three.

Whilst this uncertainty is larger than the \SI{0.2}{\kelvin} \((k=2)\) reported in \cite{lar_ap_bb_bath,cons_char_lar_ap_bb}, this study highlighted what was currently achievable. Later work will finesse the methodology and so reduce these uncertainties.

%--------------------------------------------------------
%	Conclusion
%--------------------------------------------------------

\section{\bf Conclusion}\label{sec:conc}
Rigorous assessment of the measurement environment has highlighted the practical challenges for sub-millimetre quantitative thermal imaging.

The measurement accuracy requires stricter control if characterisation over a wider temperature range is desired. For example, in semiconductor and integrated circuit regions, environments up to \SI{700}{\kelvin} will be present. This would necessitate well-defined behaviour of the optics working in close proximity to those heat sources, or a reconfiguration of the optical setup.

Preliminary assessment of the size-of-source effect indicates the severe effect it has on reliable temperature measurement. Finer aperture steps below \SI{10}{\milli\metre} may yield an improved correction.

Isolation of the thermal imager non-uniformity incurs a degree of uncertainty, but provides a desirable correction to the field-of-view of the thermal imager such that confidence in measurement across the entire temperature array was possible. The uniformity of the reference sample was a component of interest, yet may be of less concern in later work incorporating the carbon nanotube black body reference.

The \SI{640}{\milli\kelvin} \((k=2)\) combined uncertainty at \SI{320.67}{\kelvin}, this provides a sound foundation for concerted improvement. Utilising a carbon nanotube reference would provide higher emissivities and will be the focus of further study, alongside establishing traceability to the national temperature standards of ITS-90.

%--------------------------------------------------------
%	References
%--------------------------------------------------------

\section{\bf References}\label{sec:ref}
\bibliographystyle{unsrt}
\bibliography{towards_quantitative_small_scale_thermal_imaging}

\end{document}